\documentstyle[12pt]{article}
\def\be{\beta}
\def\ga{\gamma}
\def\de{\delta}
\def\ep{\epsilon}

\def\et{\eta}
\def\th{\theta}

\def\la{\lambda}

\def\ta{\tau}

\def\ph{\phi}

\def\ch{\chi}

\def\Ga{\Gamma}
\def\De{\Delta}

\def\La{\Lambda}

\def\Ps{\Psi}
\def\Om{\Omega}

\def\fr#1#2{{{#1} \over {#2}}}

\def\prt{\partial}

\def\ket#1{|{#1}\rangle}

\def\half{{\textstyle{1\over 2}}}
\def\lsim{\mathrel{\rlap{\lower4pt\hbox{\hskip1pt$\sim$}}
    \raise1pt\hbox{$<$}}}
\def\gsim{\mathrel{\rlap{\lower4pt\hbox{\hskip1pt$\sim$}}
    \raise1pt\hbox{$>$}}}

\def\X{\hat X}
\def\Y{\hat Y}
\def\Z{\hat Z}
\def\x{\hat x}
\def\y{\hat y}
\def\z{\hat z}
\def\Re{\hbox{Re}\,}
\def\Im{\hbox{Im}\,}

\newcommand{\beq}{\begin{equation}}
\newcommand{\eeq}{\end{equation}}
\newcommand{\bea}{\begin{eqnarray}}
\newcommand{\eea}{\end{eqnarray}}
\newcommand{\rf}[1]{(\ref{#1})}

\begin{document}

\title{
Bounds on CPT and Lorentz Violation from Experiments with Kaons%
\footnote{Invited talk at KAON '99, Chicago, Illinois, June 1999}
}    
\author{V.\ Alan Kosteleck\'y
\thanks{Physics Department, Indiana University, 
          Bloomington, IN 47405, U.S.A.}}
\date{IUHET 411, July 1999} 
\maketitle

\begin{abstract}
Possible signals for indirect CPT violation 
arising in experiments with neutral kaons are considered
in the context 
of a general CPT- and Lorentz-violating standard-model extension.
Certain CPT observables can depend on the meson momentum 
and exhibit sidereal variations in time.
Any leading-order CPT violation would be controlled by four parameters
that can be separately constrained in appropriate experiments.
Recent experiments bound certain combinations of these parameters 
at the level of about $10^{-20}$ GeV.

\end{abstract}

Experiments using neutral-meson oscillations
can place constraints of remarkable precision
on possible violations of CPT invariance.
For kaons,
recent results 
\cite{e773,ktevproc,cplearproc}
bound the CPT figure of merit
$r_K \equiv |m_K - m_{\overline{K}}|/m_K$
to less than a part in $10^{18}$.
Other experiments
\cite{kloe,kloeproc,gt} 
are expected to improve this bound in the near future.
Experiments with neutral-$B$ mesons
\cite{opal,delphi}
have also placed high-precision constraints
on possible CPT violation,
and the $B$ and charm factories should 
produce additional bounds
on the heavy neutral-meson systems.

A purely phenomenological treatment of possible CPT violation
in the kaon system has been known for some time 
\cite{lw}.
In this approach,
a complex phenomenological parameter $\de_K$
allowing for indirect CPT violation is introduced 
in the standard relationships between the physical meson states
and the strong-interaction eigenstates.
No information about $\de_K$ itself can be obtained
within this framework.
However,
over the past ten years
a plausible theoretical framework 
allowing the possibility of CPT violation
has been developed.
It involves the notion of 
spontaneous breaking of CPT and Lorentz symmetry
in a fundamental theory
\cite{kps},
perhaps arising at the Planck scale
from effects in a quantum theory of gravity or string theory,
and it is compatible 
both with established quantum field theory
and with present experimental constraints.
At low energies,
a general CPT- and Lorentz-violating 
standard-model extension emerges 
that preserves gauge invariance and renormalizability 
\cite{kp,cksm}
and that provides an underlying basis for the phenomenology 
of CPT violation in the kaon system.
The resulting situation is comparable to that 
for conventional CP violation,
where the nonzero value of the phenomenological parameter $\ep_K$ 
for T violation in the kaon system
can in principle be calculated 
from the usual standard model of particle physics
\cite{rs,ww}.

In this talk,
the primary interest is in the application
of the standard-model extension to CPT tests with kaons.
However,
the standard-model extension also 
provides a quantitative microscopic framework
for CPT and Lorentz violation 
that can be used to evaluate and compare a wide variety 
of other experiments
\cite{cpt98}.
These include tests
with heavy neutral-meson systems 
\cite{opal,delphi,kp,ckpv,ak},
studies of fermions in Penning traps
\cite{pennexpts,bkr,gg,hd,rm},
constraints on photon birefringence and radiative QED effects
\cite{cksm,cfj,jk,pv},
hydrogen and antihydrogen spectroscopy
\cite{antih,bkr2},
clock-comparison experiments
\cite{ccexpt,kl},
measurements of muon properties
\cite{bkl},
cosmic-ray and neutrino tests
\cite{cg},
and baryogenesis
\cite{bckp}.

Developing a plausible theoretical framework for CPT violation 
without radical revisions of established quantum field theory
is a difficult proposition
\cite{cpt98,emn}.
It is therefore perhaps to be expected that 
in the context of the standard-model extension
the parameter $\de_K$
displays features previously unexpected, 
including dependence on momentum magnitude and orientation.
The implications include,
for instance,
time variations of the measured value of $\de_K$ 
with periodicity of one sidereal (not solar) day
\cite{ak}.

First,
consider some general theoretical features 
relevant for oscillations in any neutral-meson system.
Denote generically the strong-interaction eigenstate by $P^0$,
where $P^0$ is one of $K^0$, $D^0$, $B_d^0$, $B_s^0$,
and denote the opposite-flavor antiparticle by $\overline{P^0}$.
Then a neutral-meson state is a linear combination
of the Schr\"odinger wave function for $P^0$ and $\overline{P^0}$.
The time evolution of the associated two-component object $\Ps$
state is given 
\cite{lw}
in terms of a 2$\times$2 effective hamiltonian $\La$ by 
$i\prt_t \Ps = \La \Ps$.
The physical propagating states are  
the eigenstates $P_S$ and $P_L$ of $\La$.
They have eigenvalues 
$\la_S \equiv m_S - \half i \ga_S$ and
$\la_L \equiv m_L - \half i \ga_L$,
respectively,
where $m_S$, $m_L$ are the propagating masses
and $\ga_S$, $\ga_L$ are the associated decay rates.
Flavor oscillations between $P^0$ and $\overline{P^0}$
are controlled by the off-diagonal components of $\La$,
while indirect CPT violation 
\cite{fn0}
occurs
if and only if the diagonal elements of $\La$
have a nonzero difference
$\La_{11} - \La_{22} \neq 0$.
Writing $\La$ as $\La \equiv M - \half i \Ga$,
where $M$ and $\Ga$ are hermitian,
the condition for CPT violation becomes
$\De M - \half i \De \Ga \neq 0$,
where
$\De M \equiv M_{11} - M_{22}$
and $\De \Ga \equiv \Ga_{11} - \Ga_{22}$.

A perturbative calculation 
in the general standard-model extension provides 
the dominant CPT-violating contributions to $\La$
\cite{kps}.
It turns out that the hermiticity of the perturbing hamiltonian
enforces $\De\Ga = 0$ at leading order.
The leading-order signal therefore arises 
in the difference $\De M$,
and so the standard figure of merit 
\beq
r_P \equiv \fr{|m_P - m_{\overline{P}}|}{m_P} =\fr{|\De M|}{m_P}
\label{rp}
\eeq
provides a complete description 
of the magnitude of the dominant CPT-violating effects.
An explicit expression for $\De M$ 
in terms of quantities in the standard-model extension
is known \cite{kp,ak}.
For several reasons, 
its form turns out to be relatively simple,
\beq
\De M \approx \be^\mu \De a_\mu
\quad .
\label{dem}
\eeq
Here,
$\be^\mu = \ga (1, \vec \be )$
is the four-velocity of the
meson state in the observer frame
and 
$\De a_\mu$
is a combination of CPT- and Lorentz-violating coupling constants
for the two valence quarks in the $P^0$ meson.
Note that
the oscillation experiments considered here 
provide the only known sensitivity to $\De a_\mu$.
Note also that 
the velocity dependence and the corresponding momentum dependence of $\De M$ 
is compatible with the anticipated substantial modifications
to standard physics if the CPT theorem is violated.

The experimental implications 
of momentum dependence in observables  
for CPT violation are substantial.
Effects can be classified according 
to whether they arise primarily from a dependence 
on the magnitude of the boost 
or from the variation with its direction 
\cite{ak}.
The dependence on momentum magnitude implies 
the possibility of increasing the CPT reach
by changing the meson boost 
and even the possibility of increasing sensitivity
by restricting attention to a momentum subrange in a given dataset.
The dependence on momentum direction implies 
variation of observables with the beam direction for collimated mesons,
variation with the meson angular distribution for other situations,
and sidereal effects arising from the rotation of the Earth
relative to the constant 3-vector $\De\vec a$.
In actual experiments 
the momentum and angular dependences
are frequently used to determine detector 
properties and experimental systematics,
so there is a definite risk of
cancelling or averaging away CPT-violating effects.
However,
the detection of a momentum dependence in observables
would be a unique feature of CPT violation.
There are also new possibilities for data analysis.
For instance,
measurements of an observable can be binned according to sidereal time
to search for possible time variations as the Earth rotates.

The above discussion holds for any neutral-meson system.
For definiteness,
the remainder of this talk considers the special case of kaons.
The parameter $\de_K$,
which is effectively a phase-independent quantity,
can be defined through 
the relationship between the eigenstates of the strong interaction
and those of the effective hamiltonian:
\bea
\ket{K_S} &=&
\fr{ (1 + \ep_K + \de_K) \ket{K^0}
    +(1 - \ep_K - \de_K) \ket{\overline{K^0}}  }
   { \sqrt{2( 1 + |\ep_K + \de_K|^2)}  }
\quad ,
\nonumber \\
\ket{K_L} &=&
\fr{ (1 + \ep_K - \de_K) \ket{K^0}
    -(1 - \ep_K + \de_K) \ket{\overline{K^0}}  }
   { \sqrt{2( 1 + |\ep_K - \de_K|^2)}  }
\quad .
\label{epde}
\eea
Assuming that all CP violation is small, 
$\de_K$ is in general given as 
\beq 
\de_K \approx {\De \La}/{2\De \la}
\quad ,
\label{dk}
\eeq
where 
$\De \la \equiv \la_S - \la_L$
is the eigenvalue difference of $\La$. 
In terms of the mass and decay-rate differences
$\De m \equiv m_L - m_S$
and $\De \ga \equiv \ga_S - \ga_L$,
it follows that 
$\De\la = - \De m - \half i \De \ga
= - i \De m e^{-i\hat\ph}/\sin\hat\ph$,
where $\hat\ph \equiv \tan^{-1}(2\De m/\De \ga )$.

In the context of the standard-model extension,
the above expressions show that 
a meson with velocity $\vec\be$
and corresponding boost factor $\ga$
displays CPT-violating effects given by
\beq
\de_K \approx i \sin\hat\ph ~ e^{i\hat\ph} 
\ga(\De a_0 - \vec \be \cdot \De \vec a) /\De m
\quad .
\label{dek}
\eeq
The conventional figure of merit $r_K$ becomes
\bea
r_K &\equiv & \fr {|m_K - m_{\overline{K}}|}{m_K}
\approx 
\fr{2 \De m} {m_K \sin\hat\ph} |\de_K|
\nonumber \\
&\approx & \fr { |\be^\mu\De a_\mu| }{m_K}
\quad .
\label{rk}
\eea
After substitution for the known experimental values
\cite{pdg}
for $\De m$, $m_K$, and $\sin\hat\ph$,
this gives
\beq
r_K \simeq
2\times 10^{-14} |\de_K|
\simeq 2 \left| \be^\mu \fr {\De a_\mu} {1~ \rm GeV} \right|
\quad .
\label{rknum}
\eeq
A constraint on $|\de_K|$ of about $10^{-4}$
corresponds to a limit on $|\be^\mu\De a_\mu|$
of about $10^{-18}$ GeV.

The dependence of the eigenfunctions and eigenvalues of $\La$
on $M_{11}$ and $M_{22}$ raises 
the possibility of leading-order momentum dependence in
the parameter $\ep_K$,
in the masses and decay rates
$m_S$, $m_L$, $\ga_S$, $\ga_L$,
and in various associated quantities 
such as $\De m$, $\De \ga$, $\hat\ph$.
However, 
this possible dependence is in fact absent 
because the CPT-violating contribution from 
$M_{22}$ is the negative of that from $M_{11}$,
and only $\de_K$ is sensitive to $\De M$ at leading order. 
Thus,
for example,
the usual parameter $\ep_K$ for indirect T violation 
is independent of momentum in the present framework
\cite{bell}.

The expressions obtained above can be viewed as
defined in the laboratory frame.
To exhibit the time dependence of $\de_K$ 
arising from the rotation of the Earth,
a different and nonrotating frame is useful
\cite{kl}.
A basis $(\X,\Y,\Z)$ for 
this frame can be introduced in terms 
of celestial equatorial coordinates.
The $\Z$ axis is defined as 
the rotation axis of the Earth,
while $\X$ has declination and right ascension 0$^\circ$
and $\Y$ has declination 0$^\circ$ and
right ascension $90^\circ$.
This provides a right-handed orthonormal basis
that is independent of any particular experiment.
Denote the spatial basis in the laboratory frame as 
$(\x,\y,\z)$,
where $\z$ and $\Z$ differ by a 
nonzero angle given by $\cos{\ch}=\z\cdot\Z$.
Then,
$\z$ precesses about $\Z$ with 
the Earth's sidereal frequency $\Om$.
A convenient choice of $\z$ axis 
is often along the beam direction.
If the origin of time $t=0$ is taken such that $\z(t=0)$
is in the first quadrant of the $\X$-$\Z$ plane
and if $\x$ is defined perpendicular to $\z$ 
and lies in the $\z$-$\Z$ plane for all $t$,
then a right-handed orthonormal basis
can be completed with $\y:=\z\times\x$.
It follows that
$\y$ lies in the plane of the Earth's equator
and is perpendicular to $\Z$ at all times.
Disregarding relativistic effects
due to the rotation of the Earth,
a nonrelativistic transformation
(given by Eq.\ (16) of Ref.\ \cite{kl})
provides the conversion between the two bases.

Using the above results,
one can obtain in the nonrotating frame
an expression for the parameter $\de_K$
in the general case of a kaon with three-velocity
$\vec\be = \be (\sin\th\cos\ph, \sin\th\sin\ph, \cos\th)$.
Here,
$\th$ and $\ph$ are standard spherical polar coordinates
specified in the laboratory frame about the $\z$ axis.
If $\z$ coincides with the beam axis, 
the spherical polar coordinates can be taken 
as the usual polar coordinates for a detector.
One finds
\bea
\de_K(\vec p, t) &=& 
\fr {i \sin\hat\ph ~ e^{i\hat\ph}}
{\De m} \ga(\vec p)
\Bigl[
\De a_0 + \be (\vec p) \De a_Z 
(\cos\th\cos\ch - \sin\th \cos\ph\sin\ch)
\nonumber\\
&&
\qquad \qquad \qquad 
+\be (\vec p) \left(
-\De a_X \sin\th\sin\ph 
\right . 
\nonumber\\
&&
\qquad \qquad \qquad \qquad
\left. 
+\De a_Y (\cos\th\sin\ch + \sin\th\cos\ph\cos\ch )
\right) \sin\Om t
\nonumber\\
&&
\qquad \qquad \qquad 
+\be (\vec p) \left(
\De a_X (\cos\th\sin\ch + \sin\th\cos\ph\cos\ch )
\right . 
\nonumber\\
&&
\qquad \qquad \qquad \qquad
\left. 
+\De a_Y \sin\th\sin\ph 
\right) \cos\Om t
\Bigr]
\quad ,
\label{dept}
\eea
where 
$\ga(\vec p) = \sqrt{1 + |\vec p|^2/m_K^2}$
and $\be(\vec p) = |\vec p|/m\ga(\vec p)$,
as usual.
This expression has direct implications for experiment.
For example,
the complex phase of $\de_K$ is $i \exp (i\hat\ph)$,
independent of momentum and time.
The real and imaginary parts of $\de_K$
therefore exhibit the same momentum and time dependence,
and so 
$\Re\de_K$ and $\Im\de_K$
scale proportionally when a meson is boosted.
Another property of Eq.\ \rf{dept}
is the variation of the CPT-violating effects 
with the meson boost.
For example, 
if $\De a_0 = 0$ in the laboratory frame
then there is no CPT violation for a meson at rest
but effects appear when the meson is boosted.
In contrast,
for the case where $\De \vec a = 0$ in the laboratory frame,
CPT violation is enhanced by the boost factor $\ga$ 
relative to a meson at rest.
Other implications follow from the
angular dependence in Eq.\ \rf{dept}
and from the variation of $\de_K$ with sidereal time $t$.
For example,
under some circumstances all CPT violation can average to zero
if, as usual,
neither angular separation nor time binning are performed.

The momentum and time dependence given by Eq.\ \rf{dept}
implies that the experimental setup and data-taking procedure 
affect the CPT reach.
Space restrictions here preclude consideration of 
all the different classes of scenario realized in practice.
Instead,
attention is restricted here to a single one,
typified by the E773 and KTeV experiments
\cite{e773,ktev}.
This class of experiment,
which involves highly collimated uncorrelated kaons 
having nontrivial momentum spectrum and large mean boost,
is particularly relevant here because
the KTeV collaboration announced at this conference 
the first constraints 
on the sidereal-time dependence of CPT observables 
in the kaon system
\cite{ktevproc}.
A discussion of some issues relevant to other types of experiment
can be found in Ref.\ \cite{ak}.

The KTeV experiment involves kaons with $\be\simeq 1$ 
and average boost factor $\overline\ga$ of order 100.
For this case,
$\z\cdot\Z = \cos\ch \simeq 0.6$.
In all experiments with boosted collimated kaons,
Eq.\ \rf{dept} simplifies 
because the kaon three-velocity in the laboratory frame
can be taken as  $\vec\be = (0,0,\be )$.
The expression for $\de_K$ becomes
\beq
\de_K (\vec p, t) = 
\fr {i \sin\hat\ph ~ e^{i\hat\ph}} {\De m} \ga
[ \De a_0 + \be \De a_Z \cos\ch 
+ \be \sin\ch ( \De a_Y \sin\Om t + \De a_X \cos\Om t ) ].
\label{deptktev}
\eeq
In this equation,
each of the four components of $\De a_\mu$
has momentum dependence through the boost factor $\ga$.
However,
only the coefficients of $\De a_X$ and $\De a_Y$
vary with sidereal time. 

To gain insight into the implications 
of Eq.\ \rf{deptktev},
consider first a conventional analysis
that seeks to constrain the magnitude $|\de_K|$ 
but disregards the momentum and time dependence.
Assuming the experiment is performed over an extended time period,
as is typically the case,
the relevant quantity is the time and momentum average
of Eq.\ \rf{deptktev}:
\beq
|\overline {\de_K}| = 
\fr {\sin\hat\ph } {\De m} \overline{\ga}
( \De a_0 + \overline{\be} \De a_Z \cos\ch )
\quad ,
\label{deptktevav2}
\eeq
where 
$\overline{\be}$ and $\overline{\ga}$ 
are appropriate averages of $\be$ and $\ga$,
respectively,
taken over the momentum spectrum of the data.
Substitution of the experimental quantities
and the current constraint on $|\de_K|$ from this class
of experiment
permits the extraction of a bound on a combination of
$\De a_0$ and $\De a_Z$
\cite{ak}:
\beq
|\De a_0 + 0.6 \De a_Z| \lsim 10^{-20} {\rm ~ GeV}
\quad .
\label{bound1}
\eeq 
The ratio of this to the kaon mass
compares favorably with the ratio of the kaon mass
to the Planck scale.
Note that the CPT reach of this class of experiments
is some two orders of magnitude greater than might be
inferred from the bound on $r_K$,
due to the presence of the boost factor 
$\overline{\ga} \simeq 100$.

In experiments with kaon oscillations,
the bounds obtained on $\de_K$ 
are extracted from measurements on other observables
including,
for instance,
the mass difference $\De m$,
the $K_S$ lifetime $\ta_S= 1/\ga_S$,
and the ratios $\et_{+-}$, $\et_{00}$ 
of amplitudes for $2\pi$ decays.
The latter are defined by
\bea 
\et_{+-} &\equiv &
\fr {A(K_L \to \pi^+\pi^-)} {A(K_S \to \pi^+\pi^-)}
\equiv | \et_{+-} | e^{i\ph_{+-}} 
\approx \ep + \ep^\prime
\quad ,
\nonumber \\
\et_{00} &\equiv &
\fr {A(K_L \to \pi^0\pi^0)} {A(K_S \to \pi^0\pi^0)}
\equiv | \et_{00} | e^{i\ph_{00}}
\approx \ep - 2\ep^\prime
\quad .
\label{etas}
\eea
Adopting the Wu-Yang phase convention \cite{wy},
it follows that 
$\ep \approx \ep_K - \de_K$ 
\cite{barmin,td}.
Experimentally,
it is known that $|\ep| \simeq 2\times 10^{-3}$
\cite{pdg}
and that $|\ep^\prime| \simeq 6\times 10^{-6}$
\cite{ktevcp}.
Since $\de_K$ is bounded only to about $10^{-4}$
it is acceptable at present to neglect $\ep^\prime$,
equivalent to assuming the hierarchy  
$|\ep_K| > |\de_K| > |\ep^\prime|$.
Noting that the phases of $\ep_K$ and $\de_K$ differ by $90^\circ$
\cite{buch} then gives
\bea
| \et_{+-} | e^{i\ph_{+-}}
&\approx&
| \et_{00} | e^{i\ph_{00}} 
\approx 
\ep \approx \ep_K - \de_K 
\nonumber\\
&\approx& (|\ep_K| + i |\de_K|)e^{i\hat\ph}
\quad .
\label{approx}
\eea
This implies
\bea
|\et_{+-}| & \approx & 
|\et_{00}|  \approx 
|\ep_K|(1 + O(|\de_K/\ep_K|^2)
\quad ,
\nonumber\\
\ph_{+-} &\approx & 
\ph_{00} \approx 
\hat\ph + |\de_K/\ep_K|
\quad ,
\label{approx2}
\eea
which shows that 
leading-order momentum and time dependences
in measured quantities 
appear only in the phases 
$\ph_{+-}$ and $\ph_{00}$.
The momentum and time dependences 
are absent or suppressed in other observables,
including
$|\et_{+-}|$, $|\et_{00}|$, $\ep^\prime$,
$\De m$, $\hat\ph$, and $\ta_S= 1/\ga_S$.

Substituting for $\de_K$ in 
$\ph_{+-}$ and $\ph_{00}$.
yields expressions displaying explicitly the 
time and momentum dependences:
\bea
\ph_{+-} \approx \ph_{00} &\approx &\hat\ph + 
\fr {\sin\hat\ph } {|\et_{+-}| \De m} \ga
[ \De a_0 + \be \De a_Z \cos\ch 
\nonumber\\
&& 
\qquad
+ \be \sin\ch ( \De a_Y \sin\Om t + \De a_X \cos\Om t) ] .
\label{deptktev3}
\eea
Since the coefficients of each of the four components 
$\De a_0$, $\De a_X$, $\De a_Y$, $\De a_Z$
are all distinct,
this equation shows that in principle
each component can be independently bounded
in the class of experiments involving collimated kaons 
with a nontrivial momentum spectrum.
Thus,
binning in time and fitting to sine and cosine terms 
would allow independent
constraints on $\De a_X$ and $\De a_Y$,
while a time-averaged analysis 
would permit the extraction of $\De a_0$ and $\De a_Z$ 
provided the momentum spectrum includes a significant
range of $\vec \be$.
Note,
however,
that the latter separation is unlikely to be possible
at experiments with high mean boost
because then $\be \simeq 1$ over much of the momentum range.

A constraint $A_{+-}\lsim 0.5^\circ$
on the amplitude $A_{+-}$ of time variations of the phase $\ph_{+-}$
with sidereal periodicity was announced at this conference 
\cite{ktevproc}.
Equation \rf{deptktev3} shows that $A_{+-}$ is given by
\beq
A_{+-} =  
\be\ga \fr {\sin\hat\ph \sin\ch } {|\et_{+-}| \De m} 
\sqrt{ (\De a_X)^2 + (\De a_X)^2 }
\quad .
\label{amp}
\eeq
Substitution for known quantities and for the experimental
constraint on $A_{+-}$ places the bound
\beq
\sqrt{(\De a_X)^2 + (\De a_Y)^2} \lsim 10^{-20} {\rm ~GeV}
\quad
\label{bound2}
\eeq
on the relevant parameters for CPT violation.
Like the bound \rf{bound1},
the ratio of this bound to the kaon mass
compares favorably with the ratio of the kaon mass to
the Planck scale.
Note that the bounds \rf{bound1} and \rf{bound2}
represent independent constraints on possible CPT violation.
Note also that in principle a constraint on the phase of the
sidereal variations of $\ph_{+-}$,
determined by the ratio $\De a_Y/\De a_X$,
would permit the separation of $\De a_X$ and $\De a_Y$.

The examples discussed in this talk show that
the study of momentum and time dependence in CPT observables 
is necessary to obtain the full CPT reach in a given experiment.
Additional interesting results would emerge from careful analyses
for experiments other than the ones considered here. 
Moreover,
although emphasis has been given to the kaon system,
related analyses in other neutral-meson systems
would be well worth pursuing.


\begin{thebibliography}{99}
\bibitem{fn0}
Direct CPT violation in the decay amplitudes
is neglected in this talk because it is 
expected to be unobservable
in the standard-model extension
\cite{kp}.
\end{thebibliography}
\end{document}